\documentclass[conference]{IEEEtran}
\usepackage[
  pass,
]{geometry}

\usepackage{alltt                                    
          , multirow
          , booktabs
          , listings
          , graphicx
          ,float
	,cite
          ,verbatim
         ,mathtools
	,url
	,amsmath
}
\usepackage[table]{xcolor}
\usepackage[numbers]{natbib}     
\usepackage{syntax}
\usepackage{algorithmic, algorithm}
\usepackage{enumitem}
\usepackage{threeparttable}
\usepackage{framed}
\usepackage{hhline}
\usepackage{color, colortbl}
\usepackage{balance}

\usepackage{expl3}
\ExplSyntaxOn
\newcommand\latinabbrev[1]{
  \peek_meaning:NTF . {
    #1\@}%
  { \peek_catcode:NTF a {
      #1., \@ }%
    {#1., \@}}}
\ExplSyntaxOff

\newcommand{\CASE}[1]{\STATE \textbf{case} #1\textbf{:} \begin{ALC@g}}
\newcommand{\ENDCASE}{\end{ALC@g}}

\newcommand{\DEFAULT}{\STATE \textbf{default:} \begin{ALC@g}}
\newcommand{\ENDDEFAULT}{\end{ALC@g}}
\newcommand{\DEFAULTLINE}[1]{\STATE \textbf{default:} }
\let\footnotesize\scriptsize

\newsavebox{\supbox}
\newcommand{\bsup}{\begin{lrbox}{\supbox}$\tt\scriptstyle}
\newcommand{\esup}{$\end{lrbox}{}^{\usebox{\supbox}}}
\def\eg{\latinabbrev{e.g}}
\def\ie{\latinabbrev{i.e}}

\definecolor{lightpurple}{rgb}{0.8,0.8,1}
\definecolor{codebg}{RGB}{255,255,255}
\definecolor{commentcolor}{RGB}{11,140,11}
\definecolor{yellow}{rgb}{.93,.91,.67}
\definecolor{pastel}{rgb}{.83,.83,.76}
\definecolor{green}{rgb}{.56,.74,.56}
\definecolor{lightblue}{rgb}{.68,.85,.90}
\definecolor{Gray}{gray}{0.8}

\begin{document}
%

\title{Recommending Insightful Comments for Source Code using Crowdsourced Knowledge \vspace{-.4cm}}
%
%
%
%
%

\author{\IEEEauthorblockN{$^\ast$Mohammad Masudur Rahman  ~~~ $^\ast$Chanchal K. Roy ~~~$^\dagger$Iman Keivanloo}
\IEEEauthorblockA{$^\ast$University of Saskatchewan, Canada~~  $^\dagger$Queen's University, Canada\\
\{$^\ast$masud.rahman, $^\ast$chanchal.roy\}@usask.ca, $^\dagger$iman.keivanloo@queensu.ca}
}

\maketitle

\begin{abstract}

Recently, automatic code comment generation is proposed to facilitate program comprehension. Existing code comment generation techniques focus on describing the functionality of the source code.  
However, there are other aspects such as insights about quality or issues of the code, which are overlooked by earlier approaches.
In this paper, we describe a mining approach that recommends insightful comments about the quality, deficiencies or scopes for further improvement of the source code. 
First, we conduct an exploratory study that motivates crowdsourced knowledge from
Stack Overflow discussions as a potential resource for source code comment recommendation. 
Second, based on the findings from the exploratory study, we propose a heuristic-based technique for mining insightful comments from Stack Overflow Q \& A site for source code comment recommendation. Experiments with 292 Stack Overflow code segments and 5,039 discussion comments show that our approach has a promising recall of 85.42\%.
We also conducted a complementary user study which confirms the accuracy and usefulness of the recommended comments.


\end{abstract}

\IEEEpeerreviewmaketitle

\section{Introduction}
Studies show that software maintenance can take up to 85\%--90\% of the total cost of a software product \cite{legacy,modernizing}.
The most time-consuming task during maintenance is program comprehension, and developers spend about 50\% of their time for comprehending the code to be maintained \cite{comprehend}.
Despite several studies \cite{comprehend} demonstrating the utility of code comments in comprehension, a few software projects document their code that reduces the future maintenance costs \cite{methodcomment}.
Self-documentation of the code (\eg\ descriptive identifier names) might also occasionally lead to long identifier names which reduce the readability of the code \cite{naming}.
Thus, one way to help the developers comprehend the code is to automatically generate meaningful code comments that explain the intent of the code, and there exist several techniques for this \cite{methodcomment,methodaction,methodcontext,autocomment}.
However, for gaining a deeper understanding of the code beyond its functionality, 
the developers may find neither the simple code comments (manually written or auto-generated) nor the descriptive identifier names sufficient enough.

Existing techniques generate comments for different granularities of code such as method \cite{methodcomment,methodcontext}, code segment \cite{methodaction,autocomment}, and API method call \cite{codes}.
\citet{methodcomment} propose a technique that generates the leading comment for a Java method by analyzing both method signature and identifier names used in the method body.
\citet{autocomment} propose another code comment generation technique that mines developer's description of the posted code in Q \& A site answer as the comment for a similar code fragment.
Code comments generated by existing approaches describe what the code does.
However, there are other important aspects of the code such as its quality (\eg\ maintainability) or issues (\eg\ bugs), which are ignored by earlier approaches.
In order to reach an actionable understanding (\ie\ for code reuse or change task), one needs to further analyze the code which is often a non-trivial task especially for the novice developers.
Code comments that provide insight into the quality, issues or scopes for further improvement of the source code are likely to help the developers in this regard.   

\begin{figure}[!t]
\centering
\includegraphics[width=3.5in]{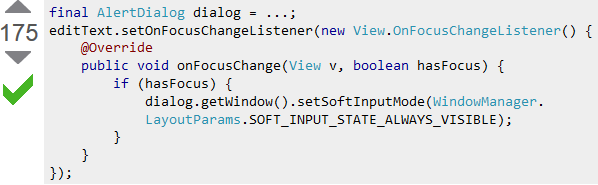}
\vspace{-.6cm}
\caption{Stack Overflow code example (taken from \cite{ce})}
\vspace{-.3cm}
\label{fig:motiv}
\end{figure}

\begin{figure}[!t]
\centering
\includegraphics[width=3.5in]{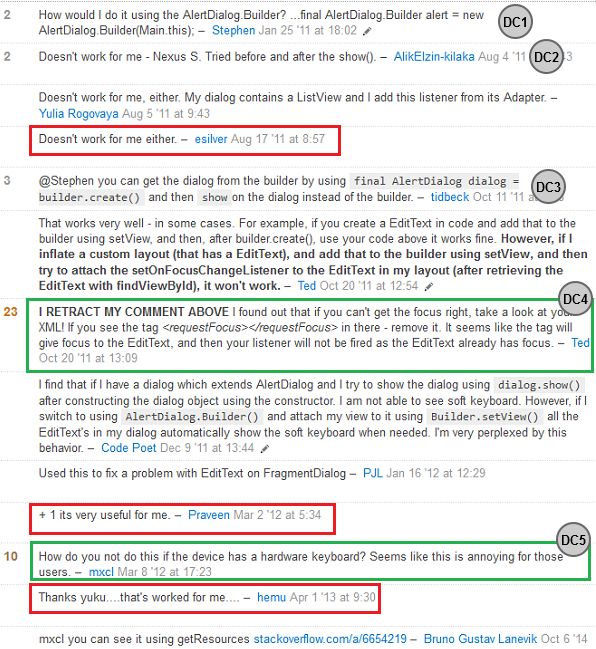}
\vspace{-.6cm}
\caption{Review comments for code example in Fig. \ref{fig:motiv}}
\vspace{-.6cm}
\label{fig:comm}
\end{figure}

In this study, we propose an approach that exploits available information on Stack Overflow to provide such additional comments.
In Stack Overflow, a posted code example (\eg\ Fig. \ref{fig:motiv}) is often followed by a series of conversations (\eg\ Fig. \ref{fig:comm}) among the participants.
Such discussion often contains useful insight that can aid a developer in analyzing a piece of code of interest. 
For instance, the code example in Fig. \ref{fig:motiv} is posted on Stack Overflow as an answer to the question-- \emph{"Android: How to show soft keyboard automatically when focus is on an EditText?"}
The code example suggests an \texttt{AlertDialog}-based technique for showing the keyboard. 
The answer is widely viewed (\ie\ 131,000 views), recognized by the community (\ie\ \emph{score}: 175), and accepted as the \emph{solution} by the person who initially posted the question.
The community also posts a number of concerns, tips and observations about the code in the form of comments. 
Apparently, one could think of these comments as the assertions based on entirely subjective viewpoints. However, such comments are constantly reviewed by other members from the community.
More importantly, the Linus's law about software bugs-- \emph{``Given enough eyeballs, all bugs are shallow\footnote{https://www.princeton.edu/$\sim$achaney/tmve/wiki100k/docs/Linus_s_Law.html}"} also perfectly applies to Stack Overflow.
Thus, such a discussion often lends itself to a high quality analysis of the posted code. 
For example, the seventh comment (\ie\ marked in green, $DC_4$, Fig. \ref{fig:comm}) in the discussion suggests an important \emph{troubleshooting tip} associated with the posted code (Fig. \ref{fig:motiv}) which is recognized by at least 23 other members from the community.
Similarly, the eleventh comment (\ie\ marked in green, $DC_5$, Fig. \ref{fig:comm}) points out an interesting concern about the code for a use case scenario, which is up-voted by at least ten other members.
Such insightful information might be invaluable in the static analysis targeting bug fixation or other maintenance
of a similar code segment (\eg\ Listing \ref{lst:segment}) from any software project (\eg\ \texttt{openmidaas-android-app}).
Unfortunately, neither manually written comments by developers nor auto-generated comments provide such information.

An automated technique for mining code comments from such resource faces a major challenge. The discussion often contains a number of trivial comments (\eg\ marked in red in Fig. \ref{fig:comm}). 
The challenge is to automatically separate the useful comments (\eg\ marked in green in Fig. \ref{fig:comm}) from the trivial ones, where the useful comments provide insights on the quality (\eg\ maintainability) or issues of the code.
Our research attempts to overcome that challenge, mines insightful comments from Stack Overflow discussions, and recommends them as code comments for a given code segment.

\begin{lstlisting}[label=lst:segment, language=java, aboveskip=10pt, belowskip=-2em, float=t, frame=bt, caption={Code segment of interest (Line 130--141 from DialogUtils.java \cite{sce})}]
final AlertDialog alertDialog = alert.create();
etName.setOnFocusChangeListener(new OnFocusChangeListener(){
		@Override
		public void onFocusChange(View arg0, boolean hasFocus) {
			if (hasFocus) { 
				alertDialog.getWindow().setSoftInputMode(
								WindowManager.LayoutParams.SOFT_INPUT_STATE_ALWAYS_VISIBLE);
			}
		}

	});
	alertDialog.show();
\end{lstlisting}

In this paper, we propose a heuristic-based technique for mining insightful comments (\ie\ discussing issues, concerns or tips) from Stack Overflow for a code segment to be comprehended or analyzed. 
We first perform an exploratory study that analyzes Stack Overflow discussions, and explores the potentials of Stack Overflow for such insightful comment recommendation.
Based on our findings from that study, we (1) collect five heuristics--\emph{popularity, relevance, comment rank, word count} and \emph{sentiment} (\ie\ polarity) associated with each of the posted comments for an Stack Overflow code example, (2) 
combine their heuristics carefully for ranking, and (3) then identify the \emph{top ranked} comments as the insightful comments.
In contrast with \citet{autocomment}, that uses the description attached to Stack Overflow code segment as comment, we explore the possibility of using follow-up discussions on Stack Overflow for insightful code comment recommendation.


An exploratory study using 9,016 questions, their accepted answers and 706 popular discussion comments from Stack Overflow shows that about 22\% of the comments are useful. They
contain meaningful insights on \emph{quality, issues} and \emph{scopes} for further improvement of the posted code which can be leveraged for automatic code comment recommendation.
We evaluate our technique in two ways.
Experiments using 292 Stack Overflow code segments from \emph{Android, Java} and \emph{C\#} domains and 5,039 discussion comments show that our mining technique has a promising recall of 85.42\% on average.
The user study involving professional developers with 85 open source code segments show that 
about 80\% of the recommended comments by our approach for those segments are found to be \emph{accurate, precise} and \emph{useful} by the participants.
Thus, we make the following contributions: 
\begin{itemize}
\item We conduct an exploratory study that demonstrates the potential of Stack Overflow for insightful code comment recommendation. 
\item We propose a heuristic-based technique for mining insightful code comments (\ie\ discussing issues, concerns, tips) from Stack Overflow for a code segment of interest.
\end{itemize}


\section{Background}\label{sec:background}

\subsection{Topic Modeling}\label{sec:topicmodel}
Topic modeling is a statistical modeling technique that retrieves underlying topic-structures of a given text corpus without the need for tags, training data or predefined taxonomies \cite{blei,barua}. It uses word frequencies and co-occurrence frequencies in the documents to develop a model of related words.
Topic modeling has been successfully used in information retrieval, software engineering and even in computer vision \cite{vision}. In software engineering, it is frequently used for topic analysis of the discussions (\eg\ mailing list) among developers \cite{barua,hot,email}, software evolution analysis \cite{topiclabel, deficient}, bug topic analysis \cite{trendy}, and even for developers' expertise analysis. 
In this paper, we use topic modeling in order to explore API topic-structures in different Stack Overflow code examples.
We use the topic modeling technique-- Latent Dirichlet Allocation (LDA) \cite{blei} for our study. LDA is a statistical topic model which considers a document  as a mixture of the retrieved topics, and considers a topic as a set of co-occurring words throughout different documents within the corpus.

\subsection{PageRank Algorithm}\label{sec:pagerank}
\emph{PageRank} algorithm by Lawrence Page and Sergey Brin is an efficient approach for ranking a list of web pages that are inter-linked with one another \cite{pagerank}. It is also widely used in other fields of research such as web spam detection, text mining, text summarization, word sense disambiguation, and natural language processing \cite{masudir}. The algorithm treats a hyper link in a web page to another website as a vote cast for that site, and it analyzes both incoming links and outgoing links of the page for the ranking.
If the web page is highly hyper-linked (\ie\ recommended) by other popular pages, it is also considered as a popular page, and vice versa. Thus, the \emph{PageRank} score of the web page can be calculated as follows:
\begin{equation}\label{eq:pagerank}
PR(A)=(1-d)+d(\frac{PR(T_1)}{C(T_1)}+... ~...+\frac{PR(T_n)}{C(T_n)})
\end{equation}
$PR(A)$ represents the \emph{PageRank} of page A, $PR(T_i)$ represents the \emph{PageRank} of pages $T_i$ that link to page A, $d$ refers to damping factor\footnote{The probability of jumping from one page to another by a visitor} that has a value between zero and one, and $C(T_i)$ is the number of outbound links on page $T_i$.

In this research, we adapt this algorithm for \emph{interaction network} of Stack Overflow comments posted for a code example, where each node denotes a comment and each edge denotes interaction between two comments.
We exploit user reference relationship (\eg\ user \emph{tidbeck} refers to user \emph{Stephen} in the fourth comment at Fig. \ref{fig:comm}) and posting sequence of comments (\eg\ $DC1\rightarrow DC2\rightarrow DC3$ at Fig. \ref{fig:comm}) for developing the comment interaction network.

\begin{figure}[!t]
\centering
\includegraphics[width=3.5in]{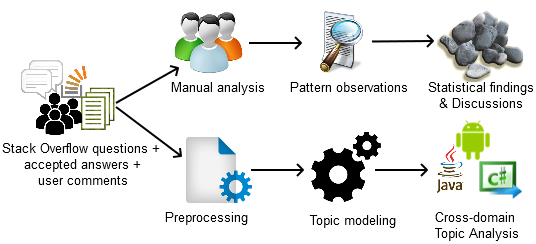}
\vspace{-.8cm}
\caption{The overview of the exploratory study}
\vspace{-.6cm}
\label{fig:explo}
\end{figure}

\section{Exploratory Study}
\label{sec:exploratory}
One of the primary goals of this research is to find out whether Stack Overflow discussions (\ie\ crowdsourced knowledge) are a meaningful source for insightful comments about code, and whether such discussions can be recommended as code comments.
We thus manually analyze 706 comments from Stack Overflow discussions in this exploratory study.
We then analyze the Stack Overflow code examples that encourage those discussions.
Fig. \ref{fig:explo} shows the schematic diagram of our study using Stack Overflow questions, their accepted answers, and discussion comments, where we perform manual analysis and cross-domain API topic agreement analysis for answering the following research questions:
\begin{itemize}
\item{\textbf{Exp-RQ$_1$:} Do the follow-up discussions from Stack Overflow contain any useful information (\eg\ insight on quality or issues of code) that is likely to aid software maintenance activities such as bug fixation or further code quality improvement?}
\item{\textbf{Exp-RQ$_2$:} Which API classes and methods are used in Stack Overflow code examples that encourage those insightful discussions?}
\end{itemize}

\begin{table}
\centering
\caption{Dataset for Manual Analysis}\label{table:manualds}
\vspace{-.2cm}
\resizebox{2.8in}{!}{%
\begin{threeparttable}
\begin{tabular}{l|c|c|c|c}
\hline
\textbf{Items}&\textbf{Java} & \textbf{Android}& \textbf{C\#} & \textbf{Total}\\ 
\hline
Questions & 98 & 81 & 103 & 282 \\
\hline
Accepted answers & 98 & 81 & 103 & 282 \\
\hline
Code segments &101 &83 & 108 & 292\\
\hline
Discussion comments &276 &161 &269 & 706\\
\hline
\end{tabular}
\end{threeparttable}
\vspace{-.2cm}
}
\vspace{-.3cm}
\end{table}

\begin{table}
\centering
\caption{Dataset for API Topic Analysis}\label{table:apids}
\vspace{-.2cm}
\resizebox{2.8in}{!}{%
\begin{threeparttable}
\begin{tabular}{l|c|c|c|c}
\hline
\textbf{Items}&\textbf{Java} & \textbf{Android}& \textbf{C\#} & \textbf{Total}\\ 
\hline
Questions & 2,932 & 2,657 & 3,427 & 9,016 \\
\hline
Code segments & 2,544 & 2,438 & 3,208 & 8,190\\
\hline
\end{tabular}
\end{threeparttable}
\vspace{-.2cm}
}
\vspace{-.5cm}
\end{table}

\subsection{Data Collection}
\label{sec:data}
We collect a total of 706 comments from the discussions of 282 accepted Stack Overflow answers from three domains-- \emph{Java, Android} and \emph{C\#} using Stack Exchange Data API\footnote{http://data.stackexchange.com/stackoverflow/queries}.
Since we manually analyze the comments for meaningful information, we choose each of them under certain restrictions-- (1) the Stack Overflow answer for which the comment is posted should contain at least ten lines of code, in order to ensure that the answer is mostly about programming, and (2) each chosen comment should be voted by at least five Stack Overflow users, in order to avoid less important comments. 
It should be noted that Stack Overflow contains millions of comments, and manually analyzing them all is impractical.
We thus collect a popular and non-trivial subset of 706 comments from a total of 5,039 comments from 282 Stack Overflow answers,
and Table \ref{table:manualds} shows the dataset for our manual analysis.

We also collect a set of 9,016 Stack Overflow questions and their accepted answers (Table \ref{table:apids}) for the study from the three domains. 
We extract code segments from the natural language text of answers using \emph{Jsoup}, a popular HTML parser.
Given that we are interested to analyze such code segments that encourage follow-up discussions, we chose the questions and their accepted answers under certain restrictions-- (1) each of the questions must be viewed at least 500 times to ensure that the question is widely viewed by the community, (2) the answer must contain one or more code segments
as a part of the solution, where each segment contains at least three lines of code, and (3) the answer must have at least ten comments by users.
In Stack Overflow, code segments are generally posted using \texttt{<code>} tags \cite{surfclipse}, and we extract the inner text of those tags as code segments.
We also automatically check the code segments using several heuristics, and discard 6\%-13\% segments as false positives from each domain, that leaves us with a total of 8,190 segments.
False positives mainly occur due to wrong content (\ie\ other than code) inside those tags.

\subsection{Answering Exp-RQ$_1$: Analysis of Discussion Comments}\label{sec:discanalysis}
To answer Exp-RQ$_1$, we analyze 706 discussion comments that are extracted during the data collection step (Section \ref{sec:data}).
In Stack Overflow, each discussion for an answer takes the form of a series of comments by the users, where the comments mostly focus on the posted code segments in the answer.
We analyze such review comments manually, and determine if they contain any meaningful information that might assist in the comprehension or other maintenance activities such as bug fixation or further quality improvement of a code segment of interest.
Our manual analysis is divided into the following sections:


\subsubsection{Classification of Comments}\label{sec:cclass}
We manually analyze 706 chosen discussion comments and their corresponding code segments, answers and questions, where 
we attempt to determine the intent behind the comments.
We particularly identify if the comments contain any mention of possible bugs (\ie\ does not work for certain cases) or limitations (\eg\ lack of portability, readability or security) that the code might face or if they contain any useful tips for improving the code. In Stack Overflow, users from different levels of expertise take part in the discussions, and such comments are often posted by them based on their work experience. Thus, the posted information is likely to assist one in troubleshooting or reusing that code. We identified several intents behind those comments, and categorize them into seven categories as follows:

\textbf{Clarification Question} $\mathbf{(C_1)}$: There exist several comments that request for more information or for confirmation about certain observations on the posted code, and finish with question marks.
We categorize them as \emph{clarification questions}.
For example, in Stack Overflow, the comment (ID: 6213963):
\begin{quote}
\emph{"Don't you miss a return when drawableMap contains the image ... without starting the fetching-thread?"}
\end{quote}
\noindent
is posted against a code segment that loads images in Android \texttt{ListView}. The comment identifies an important concern and requests for  feedback.

\textbf{Code Documentation} $\mathbf{(C_2)}$:
The comments that explain how a posted code segment works are categorized as \emph{code documentations}. 
We note that some of the discussion comments can also act as a documentation for the code segment which the discussion follows. For example, the comment (ID: 7867648):
\begin{quote}
\emph{"... look into how Intents work to understand this. It'll basically open an email application with the recipient, subject, and body already filled out. It's up to the email app to do the sending."}
\end{quote}
\noindent
is posted by the answerer to explain a code segment that uses \texttt{Intent} for sending email from an Android application.

\textbf{Tips \& Complementary Information} $\mathbf{(C_3)}$:
Given that a number of users with varying expertise view a posted answer and especially the code segments posted within it, their observations or suggestions on the code are worth noticing. Sometimes, those comments discuss useful tips (\eg\ workaround for certain issues) for the code or complementary information for further analysis and improvement of the code.
For example, in Stack Overflow, the comment (ID: 11919416):
\begin{quote}
\emph{"... as you do the mReceiver.setReceiver(null); in the onPause method, you should do the mReceiver.setReceiver(this); in the onResume method. Else you might not receive the events if your activity is resumed without being re-created."} 
\end{quote}
\noindent
is posted against the code segments that handle REST API calls within an Android application, and the comment provides a useful tip for further improvement of the code.

\begin{figure*}[!t]
\centering
\includegraphics[width=4.5in]{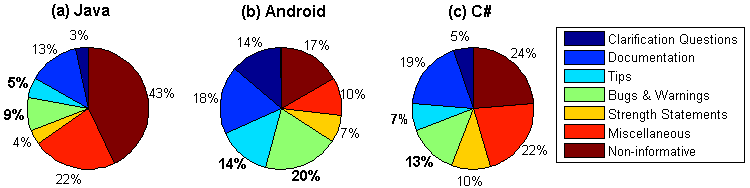}
\vspace{-.3cm}
\caption{Discussion comment statistics}
\vspace{-.5cm}
\label{fig:stats}
\end{figure*}

\textbf{Bug, Concern \& Limitation} $\mathbf{(C_4)}$: The Linus's law regarding software bugs also applies for the code segments posted on Stack Overflow. Given that a large crowd of technical users is involved in viewing and reviewing the posted answers or their code, different aspects of a code segment are explored and the discussion often points out possible bugs, errors, warnings or important concerns associated with the code. Such information is of utmost importance to other users (\eg\ developers) troubleshooting or analyzing that code. For example, one user posted the comment (ID: 10747399):
\begin{quote}
\emph{"This code does not work properly. Some '0' characters becomes missing in the generated string. I don't know why, but that's the case."} 
\end{quote}
\noindent
after reviewing an Android code segment that encrypts a string, and the comment clearly reports a bug within the code.

\textbf{Strength Statement} $\mathbf{(C_5)}$: These comments generally contain praising words about the posted code, its elegance or strength, and how it does help others in their works. 
For example, this comment (ID: 149203):
\begin{quote}
\emph{"Best damn piece of code I have seen :), Just solved a million problems in my project :)"}
\end{quote}
\noindent
is posted for a code segment that deals with dynamic \texttt{LINQ} and \texttt{IEnumerable} in C\#.
These comments also often contain trivial thank statements.


\textbf{Miscellaneous} $\mathbf{(C_6)}$: There exist comments whose intentions cannot be easily explained. They do not fall into any of the above categories. We call them \emph{miscellaneous} comments. During our analysis, we found a major part of the discussion comments belong to this genre.
For example, the comment (ID: 2903068):
\begin{quote}
\emph{"You should really cite your references. This algorithm was invented by Robert Floyd in the '60s, It's known as Floyd's cycle-finding algorithm, aka. The Tortoise and Hare Algorithm."}
\end{quote}
\noindent
is posted against a code segment that partially implements \emph{Floyd's cycle-finding algorithm} for detecting loops in linked list.
The comment emphasizes on the citation of the algorithm used in the code. Other miscellaneous comments express users' frustration about particular technology or programming language, social humour and technology history.

\textbf{Non-informative Comments} $\mathbf{(C_7)}$:
During manual analysis, we also note that a significant part of our chosen comments is not related to code although the answers contain code segments. In Stack Overflow, users sometimes post code segments in order to answer different conceptual questions where source code might not be an essential part.
In such cases, the follow-up discussion might focus on the overall answer rather than the code segments. We discard such comments from our further analysis, and categorize them as \emph{non-informative} comments.

\subsubsection{Statistics for Comment Classes}
We manually label each of the 706 comments into seven categories, which are later used as the \emph{gold comments} \cite{expds} for evaluation (Section \ref{sec:experiment}).
Fig. \ref{fig:stats} shows the fraction of different categories of popular discussion comments that follow the code segments or programming answers from the three domains- \emph{Java, Android} and \emph{C\#}.
We are particularly interested about these two categories--\emph{bugs} ($C_4$) and \emph{tips} ($C_3$) as they contain non-trivial information and the users making such comments are likely to possess a certain level of expertise on the posted code.
We observe that each of the programming domains contains a significant amount of such comments which is promising. 
For example, 20\% of \emph{C\#} comments and 14\% for \emph{Java} comments belong to these categories.
On the other hand, \emph{Android} has the maximum of 34\% comments that contain identified \emph{bugs} and improvement or troubleshooting \emph{tips} for the posted code.
One possible explanation could be that \emph{Android} is relatively less mature than the other two platforms, and thus, it encourages more discussions about bugs and further improvement. 
Such findings actually underpin our intuition about Stack Overflow discussion comments.
Although the comment classification is a bit subjective (\ie\ might face reproducibility issue), and the dataset analyzed is not large, 
the findings are still convincing enough to demonstrate that Stack Overflow comments are a potential source for insightful information about the code.

Our analysis in Exp-RQ$_1$ shows that a significant fraction of the comments posted on Stack Overflow contain useful (\ie\ insightful) information, 
and such comments are likely to aid different software maintenance activities such as bug fixation or code quality improvement.

\subsection{Answering Exp-RQ$_2$: Automated API Analysis}
Since our manual analysis suggests that Stack Overflow discussions contain meaningful information for software maintenance tasks, we are interested to investigate the program artifacts in the answers that mostly encourage those discussions.  
To answer Exp-RQ$_2$, in this section, we analyze the code segments from Stack Overflow answer texts using topic modeling, and perform cross-domain topic-agreement analysis among three domains-- \emph{Java, Android} and \emph{C\#}. 
Our semi-automated analysis has the following steps:

\subsubsection{Preprocessing} In this step, we analyze each of the code segments from the texts of 9,016 Stack Overflow answers (Table \ref{table:apids}), where we extract different API elements (\eg\ \emph{class name, method name}) from the code. We decompose dotted tokens (\eg\ \texttt{java.util.HashMap}) into individual tokens, and exploit \emph{camel-case notations} for extracting the API elements.
It should be noted that we do not include the frequent programming keywords (\eg\ \texttt{if,for}) or punctuation tokens (\eg\ \texttt{([?;}) in the list. Thus, the \emph{API corpus} of each of the three programming domains contains a list of files containing API class and method names used in the corresponding accepted answers from that domain.

\subsubsection{Performing Topic Modeling}
In order to identify API topic trends in the code segments from each programming domain, we perform topic modeling on the corresponding \emph{API corpus}. We make use of \emph{Mallet} \cite{mallet}, an LDA-based popular topic modeling tool, for the task. We extract a maximum of 150 API topics given that a large number for topics, $K$, helps identify specified or recognizable topics \cite{deficient}. We thus chose the hyper parameters-- $\alpha, \beta$ for the model carefully. We consider a symmetric version of $\alpha$, probability distribution of topics over the corpus, where all topic distributions sum to one and each distribution equals to $1/K$. A lower value of $\alpha$ indicates that the model assumes that each document in the corpus focuses on a few number of topics (\ie\ one or two) \cite{blei}. We also chose a lower value of 0.006 for $\beta$, distribution of topics over words from the corpus. A lower value of $\beta$ indicates that the model assumes that each of the topics can be represented using a few words \cite{blei,barua}. In our case, the tool automatically optimizes the topic words, and each of the 150 topics is represented using only six words. We extract such 150 API topics for each of the three programming domains under study.

\subsubsection{Cross-domain API Topic Analysis}
Once topics are extracted, we filter out the topics with a $\beta$ value less than our \emph{adopted threshold} (\ie\ 0.006) which leaves us with 26 Java API topics, 43 Android API topics, and 44 C\# API topics.
We consider the top five API topics for each of the documents in the corpus, determine \emph{topic-document-frequency}, and sort the topics according to their frequencies.
Due to space limitation, the ranked topics are available for download elsewhere\footnote{http://www.usask.ca/$\sim$mor543/codeinsight/topics}. 
We then  investigate if there exists any regular pattern among the topics from different domains.
We first assign a label to each of the topics by analyzing the API names and occasionally the associated code examples \cite{topiclabel}. We interestingly notice that several topic labels across the domains are semantically related, and we thus group them into a single category. 
Finally, we found five such categories-- (1) \emph{List \& Collections}, (2) \emph{I/O Operations}, (3) \emph{String Operations}, (4) \emph{Date \& Time}, and (5) \emph{Integer}.  
We found that among those five categories, \emph{Collections} and \emph{I/O} APIs are
commonly used in the code segments across all three domains that encourage follow-up discussions.
The other two API categories associated with string operations, date and time are dominant in \emph{Java} and \emph{C\#} domains but not in Android domain.
We also found that the API for parsing integer numbers is dominant in the code segments that attract discussions especially in \emph{Java} and \emph{Android} domains.
Such findings are particularly important since they partially explain the characteristics of the artifact (\ie\ API) that might trigger the discussions of our interest at Stack Overflow.
Besides, they might benefit API-specific comment mining techniques towards improved API documentation.
 
Our analysis in Exp-RQ$_2$ shows that several APIs  are used in the code segments that encourage follow-up discussions at Stack Overflow. However, five of them are not only frequent but also common across the three domains under study.


\begin{figure}[!t]
\centering
\includegraphics[width=3.5in]{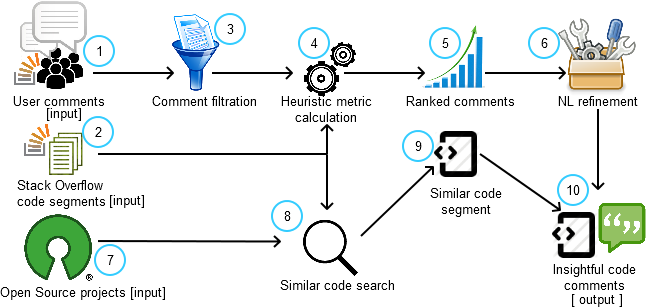}
\vspace{-.4cm}
\caption{Schematic diagram of the proposed technique}
\vspace{-.6cm}
\label{fig:sysdiag}
\end{figure}

\section{CodeInsight: Mining Insightful Code Comments for Source Code}
\label{sec:model}
Fig. \ref{fig:sysdiag} shows the schematic diagram of our proposed technique for mining insightful comments from Stack Overflow for a given code segment.
The exploratory study (Section \ref{sec:discanalysis}) analyzes potential of discussion comments from Stack Overflow, and suggests a classification for them that helps us focus on certain comment categories. 
Since we are interested about the insightful code comments discussing bugs or improvement tips for code,
a technique is required for automatically mining the discussion comments of corresponding categories--$C_4$ (\ie\ \emph{bugs}) and $C_3$ (\ie\ \emph{tips}).
We discard insignificant (\ie\ non-popular) comments from the discussion using \emph{vote count threshold} (Step 3, Fig. \ref{fig:sysdiag}), apply a set of five heuristics for ranking, and then extract the top-ranked comments (Step 4, 5). 
Finally, we refine the extracted comments using natural language processing tools (Step 6) so that they can be recommended as the code comments for a similar code segment (Step 7, 8, 9) to be analyzed for maintenance.

\subsection{Proposed Heuristics}\label{sec:heuristics}
Since our preliminary investigation shows that the type of comments (\eg\ \emph{bugs (C4)}) generally does not correlate with its linguistic characteristics such as \emph{sentence count} or \emph{readability}, we carefully choose and apply a list of five heuristics for mining the important comments. 
While \emph{popularity} and \emph{relevance} of a candidate comment with respect to the posted code are important aspects, we also heuristically capture \emph{comment size}, \emph{impact} of that comment in the discussion and the \emph{sentiment} (\eg\ positive or negative) expressed in the comment texts for identifying insightful comments as follows:

\textbf{Popularity (P)}: In Stack Overflow, comments in the follow-up discussion for a posted answer are often subjectively evaluated by other users. If a comment adds something useful to the answer, it gets up-votes, otherwise, it is marked as a trivial or non-productive comment.
In our research, we consider \emph{popularity} of comment as a heuristic for comment ranking, and we use \emph{up-vote count} as an estimate of its \emph{popularity} within the discussion.

\textbf{Word Count (WC)}: During manual analysis, we note that follow-up discussions in Stack Overflow might contain several comments that are too short to contain any meaningful information. For example, the showcase discussion (Fig. \ref{fig:comm}) contain three such comments marked in red.
We thus consider \emph{word count} as a heuristic for comment ranking, and discard such trivial comments during ranking.

\textbf{Relevance (R)}: In a follow-up discussion, there might be several comments which either apply to several code segments or do not refer to any of the code segments at all from the posted answer. These comments are not suitable for our purpose, and thus we consider \emph{relevance} of comment as another important heuristic for comment ranking. In the discussion comments, Stack Overflow users often refer to different code elements such as \emph{class names, method names} or \emph{package names} or different programming concepts (\eg\ list, sorting) from the posted code segments of an answer.
Thus lexical similarity between the code and a candidate comment is a potential estimate for relevance of that comment to the code. We use \emph{cosine similarity} measure for determining the lexical similarity \cite{surfclipse}, where we apply different preprocessing (\eg\ stop word and punctuation removal, word splitting) both on the code segments and the comments.
It should be noted that we avoid stemming for both items due to their heterogeneous nature (\eg\ code and natural language) of textual content.

\begin{figure}[!t]
\centering
\includegraphics[width=1.5in]{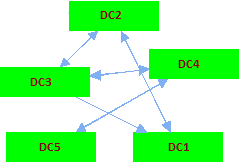}
\vspace{-.2cm}
\caption{Comment interaction network in Fig. \ref{fig:comm}}
\vspace{-.6cm}
\label{fig:inet}
\end{figure}

\textbf{Comment Rank (CR)}: In the discussion comments, Stack Overflow users sometimes refer to other users participating in the discussion using \emph{"@user"} notation. For instance, in showcase example (Fig. \ref{fig:comm}), user \emph{tidbeck} refers to user \emph{Stephen} using word-- \emph{@Stephen} at the beginning of the fifth (\ie\ $DC_3$) comment.
Such reference indicates that not only this comment is a response to the previous comment (\ie\ first comment, $DC_1$) posted by \emph{Stephen} but also the previous comment is a stimulus for the discussion. We identify such stimulating and stimulated comments in the discussion, and develop a \emph{comment interaction network}.
We also use the sequence of comments based on their \emph{comment id} in developing the interaction network, and in this case, the baseline heuristic is that a comment not only does get influenced by the immediate previous comment in the conversation but also it influences the following comment. 
For example, Fig. \ref{fig:inet} shows the interaction network for the comments (\ie\ filtered by \emph{vote count}) in Fig. \ref{fig:comm}, where each node refers to a distinct comment (\ie\ labeled with comment ID) and each edge denotes an interaction between the two comments.
We apply \emph{PageRank algorithm} (Section \ref{sec:pagerank}) on this network to recursively calculate the \emph{score} for each node by analyzing not only its connectivity in the network but also the scores of the connected nodes.
The algorithm runs until the scores for each of the nodes converge, and such a score can be considered as an estimate of relative importance of a node in the network.
We consider that score as \emph{comment rank} (\ie\ analogous to \emph{page rank} for web pages \cite{pagerank}), and apply it as a heuristic for comment ranking.

\textbf{Sentiment (S)}: From the exploratory study (Section \ref{sec:discanalysis}), we notice that discussion comments that point out possible bugs and warnings ($C_4$) or troubleshooting tips ($C_3$)  for the posted code are often associated with negative emotions such as frustration, annoyance and disappointment. On the other hand, comments that contain appreciation about the posted code emit positive emotions such as gratitude or thankfulness. Since we are interested in mining the comments containing bugs, concerns and tips for improving the code, sentiment expressed in the comment texts is a potential heuristic for separating them from rest of the comments in the discussion \cite{emotion}.  
We use a state-of-the-art sentiment analysis tool-- \emph{Stanford Sentiment Analyzer} \cite{sentiment} for determining sentiment of each of the comments. 
The tool returns a sentiment score on the scale from ``0" (\ie\ very negative) to ``4" (\ie\ very positive) for a sentence, and we adapt the scale so that it ranges from ``-2" (\ie\ very negative) to ``+2" (\ie\ very positive) and ``0" denotes neutral sentiment.
We parse each of the sentences from a candidate comment, sum up their sentiment scores, and then use it for the ranking of the discussion comments.

\subsection{Heuristic-based Comment Ranking \& Selection}
Once the discussion comments are collected from Stack Overflow, less important (\ie\ non-popular) comments are filtered out.
In particular, we solve the above problem by
considering only such comments which have scored at least one up-vote to date and discarding the rest from further analysis. 
This filtering step helps us move on with a moderate list containing relatively popular and relatively important comments.  
We then collect each of the five heuristics-- \emph{popularity, relevance, comment rank, word count} and \emph{sentiment} for those comments, and develop separate ranked lists for each of the heuristics.
It should be noted that the discussion comments are ranked in ascending order for sentiment heuristic (\ie\ negative sentiment associates with bugs) whereas they are ranked in the opposite order for the rest four heuristics. 

Since each individual heuristic captures a distinct aspect of salience for the comments, top comments in each of those five rankings are potential candidates for final selection. 
We choose \emph{top five} comments from each ranking, and determine the frequency for each of the comments in those top ranks (adapted from \citet{strathcona}).
We then choose the top $K=3$ comments based on their frequencies as the insightful comments.
Such ranking and selection of comments ensure that the extracted comments are not only \emph{popular} among the users but also \emph{non-trivial}, \emph{stimulating} for discussion and \emph{relevant} to the posted code.
More interestingly, they point out important  issues or concerns in the code in terms of negative \emph{sentiment}.
For example, our heuristic algorithm extracts $DC_4, DC_5$ and $DC_1$ from the discussion in Fig. \ref{fig:comm} as the insightful comments, where the first comment, $DC_4$, provides an important tip for troubleshooting, and the second one, $DC_5$, reports an important concern about the posted code.  

\subsection{Comment Text Refinement}
Discussion comments from Stack Overflow often contain personal pronouns (\eg\ \emph{"I", "you",} etc.) and possessive pronouns (\eg\ \emph{"mine", "yours",} etc.). As suggested by existing studies \cite{autocomment}, such words contribute no value in a comment for code, and we thus replace them with more objective words. For example, the comment--\emph{"You should not use the IV like this. For a given two messages, they should not have been encrypted with the same Key and same IV"} is transformed into \emph{"One should not use the IV like this. For a given two messages, they should not have been encrypted with the same Key and same IV"}. We use Stanford CoreNLP POS tagging \cite{sentiment} for identifying such pronouns, and replace them selectively in order to make the comments formal.
We also remove user reference information (\eg\ \emph{@Stephen}), replace numbers (\eg\ \emph{"3"}) with numerical words (\eg\ \emph{"three"}), and transform the spoken words (\eg\ \emph{"can't", "doesn't"}) into formal words (\eg\ \emph{"can not","does not"}) to make the comments useful.

\begin{table*}
\centering
\caption{Experimental Results for Different Heuristics}\label{table:component}
\vspace{-.2cm}
\resizebox{7in}{!}{%

\begin{threeparttable}
\begin{tabular}{l|l||c|c||c|c||c|c||c|c}
\hline

\multirow{2}{*}{\textbf{Heuristics Used}} & \multirow{2}{*}{\textbf{Metric}} & \multicolumn{2}{c||}{\textbf{Java}} & \multicolumn{2}{c||}{\textbf{Android}} & \multicolumn{2}{c||}{\textbf{C\#}} &  \multicolumn{2}{c}{\textbf{Average}}\\
\hhline{~~--------}

& & $\mathbf{C_3}\tnote{1}$ & $\mathbf{C_4}\tnote{2}$ & $\mathbf{C_3}$ & $\mathbf{C_4}$ & $\mathbf{C_3}$ & $\mathbf{C_4}$ & $\mathbf{C_3}$ & $\mathbf{C_4}$ \\
\hline
\hline
\multirow{3}{*}{\{Popularity (P)\}}& CE\tnote{3} & 09(15) & 16(24) & 22(23)& 28(32)& 17(19)& 25(36)&-- &-- \\
\hhline{~---------}
& Recall &60.00\% &66.67\% & 95.65\%& 87.50\%&89.47\% &69.44\% & 81.71\% &74.53\% \\
\hhline{~---------}
& MRR\tnote{4} &0.44 &\textbf{0.50} & 0.59& \textbf{0.57}&0.47 &\textbf{0.56} &0.50 &\textbf{0.54}  \\
\hline
\multirow{3}{*}{\{Popularity (P), Relevance (R)\}}& CE & 09(15) & 16(24) & 22(23)& 28(32)& 16(19)& 29(36)&-- & --\\
\hhline{~---------}
& Recall &60.00\% &66.67\% & 95.65\%& 87.50\%&84.21\% &80.56\% & 79.95\% &78.24\% \\
\hhline{~---------}
& MRR &0.44 &0.44 & 0.64& 0.54&0.50 &0.48 &0.53 & 0.49 \\
\hline
\multirow{2}{*}{\{Popularity (P), Relevance (R),}& CE & 10(15) & 17(24) & 22(23)& 27(32)& 16(19)& 28(36)&-- & --\\
\hhline{~---------}
& Recall &\textbf{66.67}\% &70.83\% & 95.65\%& 84.38\%&84.21\% &77.78\% & 82.18\% &77.66\% \\
\hhline{~---------}
Comment Rank (CR)\}& MRR &\textbf{0.60} &0.29 & \textbf{0.73}& 0.56&0.44 &0.50 &\textbf{0.59} &0.45  \\
\hline
\multirow{2}{*}{\{Popularity (P), Relevance (R),}& CE & 07(15) & 18(24) & 22(23)& 28(32)& 17(19)& 30(36)&-- & --\\
\hhline{~---------}
& Recall &46.67\% &75.00\% & 95.65\%& 87.50\%&89.47\% &83.33\% & 77.26\% &81.94\% \\
\hhline{~---------}
Comment Rank (CR), Word Count (WC)\} & MRR &0.57 &0.33 & 0.59& 0.50&\textbf{0.53} &0.47 &0.56 & 0.43 \\
\hline
\multirow{2}{*}{\{Popularity (P), Comment Rank (CR),}& CE & 09(15) & 19(24) & 22(23)& 31(32)& 18(19)& 31(36)&-- & --\\
\hhline{~---------}
& Recall &60.00\% &\textbf{79.16}\% & \textbf{95.65}\%& \textbf{96.88}\%&\textbf{94.74}\% &\textbf{86.11}\% & \textbf{83.46}\%& \textbf{87.38}\%\\
\hhline{~---------}
 Relevance (R), Word Count (WC), Sentiment (S)\}& MRR &0.44 &0.32 & 0.55& 0.52&0.33 &0.45&0.44 & 0.43 \\
\hline

\end{tabular}
\centering
 $^1$Comments containing improvement tips,  $^2$Comments containing bugs and warnings,  $^3$Comments retrieved, $^4$Mean Reciprocal Rank
\end{threeparttable}
\vspace{-.4cm}
}
\vspace{-.6cm}
\end{table*}

\section{Evaluation of CodeInsight}
\label{sec:experiment}
We conduct two experiments-- an empirical study and a user study for evaluating our proposed technique for insightful code comment generation from Stack Overflow for a given code segment. The goal is to evaluate both the performance of the comment ranking algorithm and the quality of the recommended comments \cite{autocomment}.
We thus attempt to answer the following three research questions using those experiments:
\begin{itemize}
\item \textbf{RQ1}: How effective the proposed technique is in retrieving the comments that discuss bugs, concerns and tips for improvement in the posted code?
\item \textbf{RQ2}: Are the recommended comments \emph{accurate, precise}  and \emph{concise} in describing the potential issues or troubleshooting tips for the target code segment?
\item \textbf{RQ3}: Are the recommended comments \emph{useful} for static analysis involving maintenance of the target code?
\end{itemize}

\subsection{Evaluation of Comment Ranking Algorithm}\label{sec:empirical}
In the first experiment, we evaluate our ranking algorithm that ranks a list of discussion comments from Stack Overflow based on the proposed heuristics (Section \ref{sec:heuristics}), and recommends the three most insightful comments for a code segment to be analyzed or comprehended.
We use 5,039 discussion comments targeting 292 code segments, and 706 gold comments (\ie\ manually labeled) (Table \ref{table:manualds})
from three popular domains-- \emph{Java, Android} and \emph{C\#} for this experiment.
Qualitative analysis on the gold comments \cite{expds} can be found in Section \ref{sec:discanalysis}.
Our algorithm successfully retrieves 130 out of 149 comments ($C_3$ and $C_4$) discussing \emph{bugs, warnings} and \emph{tips} about the posted code from the total of 5,039 Stack Overflow comments.
We use two performance metrics--\emph{recall} and \emph{mean reciprocal rank} (MRR) for the evaluation. \emph{Recall} denotes the fraction of the gold comments that is retrieved by a technique whereas \emph{reciprocal rank} refers to the multiplicative inverse of the first relevant item's position in the ranked list.
\emph{Mean reciprocal rank} averages such measure for all trials. It should be noted that precision is not calculated as a performance metric since we recommend only three comments at once.

From Table \ref{table:component},  we note how each of the five heuristics-- \emph{popularity, relevance, comment rank, word count} and \emph{sentiment} contributes to the final ranking and selection of the discussion comments. We incrementally add different heuristics to the ranking algorithm, and analyze their performance for avoiding any potential suboptimal set of heuristics.
We note that \emph{popularity} heuristic dominates others to a large extent, and this occurs probably due to our design choice for the dataset. 
We exploit mass evaluation for the comments by a large technical crowd in order to avoid subjective bias, and chose a popular (\ie\ up-voted) subset of all comments for manual analysis (Section \ref{sec:discanalysis}) and experiment. Although the \emph{popularity} heuristic performs considerably well in mining important \emph{tips} (\ie\ \emph{recall} 81.71\%), 
it does not perform equally (\ie\ \emph{recall} 74.53\%) for \emph{bugs} and \emph{warnings}. We thus add more heuristics gradually, and achieved a global maximum in terms of \emph{recall} for both comment types when all five heuristics are combined.
It should be noted that \emph{mean reciprocal ranks} do not show similar behaviour, and they mostly remain comparable.
In short, our comment ranking algorithm retrieves $C_3$ comments (\eg\ \emph{tips}) and $C_4$ comments (\eg\ \emph{issues} or \emph{concerns})
with a promising \emph{recall} of 85.42\% on average. 
Although the dataset (\ie\ 5K comments) used is not large, the findings clearly demonstrate the potential of our ranking technique.
Such findings also answer our first research question, RQ$_1$.

\begin{table*}
\centering

\caption{Developer Judgement on Recommended Comments}\label{table:likert}
\vspace{-.2cm}
\resizebox{6.5in}{!}{%
\begin{threeparttable}
\begin{tabular}{l||c|c|c|c||c|c|c|c||c|c|c|c||c|c|c|c}
\hline

& \multicolumn{4}{c||}{\textbf{Java}} & \multicolumn{4}{c||}{\textbf{Android}}& \multicolumn{4}{c||}{\textbf{C\#}} & \multicolumn{4}{c}{\textbf{Average}}\\ 
\hline
\textbf{Responses} & \textbf{Ac}& \textbf{Pr}& \textbf{Co}& \textbf{Us} & \textbf{Ac}& \textbf{Pr}& \textbf{Co}& \textbf{Us} & \textbf{Ac}& \textbf{Pr}& \textbf{Co}& \textbf{Us} & \textbf{Ac}& \textbf{Pr}& \textbf{Co}& \textbf{Us}\\ 
\hline
\hline
Strongly Agree & 13& 11& 8& 12 &20 &16 & 16& 18&11 &10 &12 &11& \multirow{2}{*}{82.50\%} & \multirow{2}{*}{80.83\%} & \multirow{2}{*}{78.33\%} & \multirow{2}{*}{79.17\%}\\
\hhline{-------------~~~~}
Agree &4 &7 &8 & 5 &11 &11 &12 &13 & 6&7 &5 &4 & & & & \\
\hline
Neutral &2 &0 &3 &1  &2 &7 &7 &4 &1 &1 &1 & 3 &6.67\% & 7.50\%&12.50\%  & 10.00\%\\
\hline
Disagree &0 &1 &0 &1  &4 &4 &4 &3 &2 &2 &2 &2 & \multirow{2}{*}{10.83\%}  & \multirow{2}{*}{10.00\%}  &  \multirow{2}{*}{9.17\%} & \multirow{2}{*}{10.83\%}\\
\hhline{-------------~~~~}
Strongly Disagree &1 &1 &1 &1  &3 & 2& 1& 2& 0& 0& 0& 0 & & & &\\
\hline
\hline
Total &20 & 20&20 &20  &40 &40 &40 & 40&20 &20 &20 &20 \\
\hline
\end{tabular}
\centering
\textbf{Ac}=Accurate, \textbf{Pr}=Precise, \textbf{Co}=Concise, and \textbf{Us}=Useful 
\end{threeparttable}
}
\vspace{-.6cm}
\end{table*}


\subsection{Evaluation of Recommended Comments: A User Study}
Although the empirical evaluation clearly shows the high potential of our proposed technique, we also wanted to see whether the developers do like the technique and find the recommended comments useful.
This leads to our second evaluation, and we conducted a user study involving four professional developers as well. 
In collaborative software development, software developers often comprehend or analyze the code written by peers in the form of code reviews or code reuse. We attempt to determine if the professional developers, \ie\, study participants, find our recommended comments \emph{accurate, precise, concise} or \emph{useful} for analyzing the code segments that are similar to Stack Overflow code segments but are taken from open source projects \cite{autocomment}.

\textbf{Dataset Preparation for User Study}: We conduct a case study with 82 open source projects for collecting dataset for the user study. Our exploratory study (Section \ref{sec:discanalysis}) and empirical evaluation (Section \ref{sec:empirical}) involve 292 Stack Overflow code segments, and in this case, we locate similar segments in open source projects.
We exploit \emph{GitHub code search} for finding similar code since it is a preferable choice for our task. 
GitHub hosts a large collection of open source projects, and its \emph{code search feature} extracts results from hundreds of projects. 
Thus, it has a greater chance of retrieving the code segments of interest, and existing studies \cite{surfexample} also show such findings. 
We also found that code search using appropriate keywords followed by a careful manual analysis of the top results can actually locate the code segments of interest.

We perform an exhaustive search using GitHub search, and locate 85 code segments similar to Stack Overflow code segments in the open source projects. 
We found that most of them are either directly copied (\ie\ Type 1 clone) or slightly modified (\ie\ Type 2 clone) before posting to Stack Overflow. Thus, they are suitable candidates for the user study that evaluates the quality of the comments from Stack Overflow recommended for similar code segments.

\textbf{Study Participants}: In our user study, we involve four professional developers from two reputed software development companies specialized in web technology and mobile applications with 5-10 years of experience.
The developers have professional experience on each of the three programming domains that ranges from 1 to 2.5 years for standard \emph{Java}, 1.5--5 years for \emph{Android} and 1.5--3.5 years for \emph{C\#}.
Such experience makes them suitable candidates for our study.

\vspace{-.1cm}
\textbf{Study Setup}: In the study, each of the participants worked with 20 code segments (\ie\ 5 for \emph{Java}, 10 for \emph{Android} and 5 for \emph{C\#}) and spent about 2 hours on average.
The code segments are randomly chosen from the collection of 85 open source code segments. 
We provide top three comments by our technique for each of those code segments (along with their technical problem context), and ask the participants to evaluate those comments.
The goal is to determine if such comments can provide any meaningful information beyond simple explanation, \ie\ can provide insights about quality, issues or tips for further improvement of the code. 
We ask each of the participants to report their responses about the quality of the comments in terms of \emph{accuracy, preciseness, conciseness} and \emph{usefulness} using a five-point Likert scale.
We collect 80 responses against those four quality aspects from each of the participants, and this provides a total of 320 data points for evaluation. 
Thus, we collect sufficient data points, and also partially handle the threat involving small number of participants.   
However, adding more participants and using more data points would surely further strengthen the findings.  

\textbf{Study Results \& Discussions}: Table \ref{table:likert} summarizes our findings from the user study. 
We note that each of the participants found most of the recommended comments \emph{accurate}, \emph{precise}, \emph{concise} and \emph{useful} 
from each of the three domains.
In particular, around 83\% of the comments for \emph{Java} and \emph{C\#} code segments are found \emph{salient} whereas that statistic is 73.13\% for \emph{Android} code segments.
We also note that around 7\%-14\% of the comments are marked as irrelevant or non-informative whereas the participants remained undecided for 7\%-12\% of the comments for different domains.
We thus collect observations and suggestions from each of the participants, and attempt to better explain such findings.
First, they encountered a few comments which hint about certain issues in the code, but they were expecting more details (\ie\ not \emph{precise}). 
Second, one of the participants encountered three comments that refer to certain identifier names which were not in the code segment (\ie\ identifiers changed before posting). Thus, the participant found those comments \emph{inaccurate}.
Third, according to another participant, there were one comment that poses silly arguments which are not helpful for program analysis (\ie\ not \emph{useful}).
In order to address such subtle issues with Stack Overflow comments, more sophisticated filtration is a possible choice which we consider as a future work. 

Thus, according to our conducted user study, above 80\% of the recommended comments are found \emph{accurate} and \emph{precise} and about 79\% of them are found \emph{concise} and \emph{useful} by the participants.
This clearly answers RQ$_2$ and RQ$_3$, and suggests high quality of the recommended comments.

\section{Threats to Validity}
\label{sec:threats}
We identify the following threats to the validity of our study. 
First, the dataset (especially discussion comments) used for empirical evaluation (Section \ref{sec:empirical}) is limited which someone could consider as a threat.
However, such data are carefully extracted from a larger dataset based on certain well-defined restrictions (Section \ref{sec:data}). The goal is to analyze discussion comments associated with code segments, and thus, we discard a major fraction of Stack Overflow comments that are not associated with the code, \ie\ accepted answer does not contain a code example.
Besides, we had to manually analyze those comments in the exploratory study, and analyzing a large amount of comments is impractical.
Thus, the dataset might be sufficient enough to validate our findings.

Second, we only involved four professional developers in the user study. While more participants could have been useful, we should note that the empirical evaluation clearly shows the high potential of the proposed technique and 
that additional observations with the professional developers only confirm the quality of the recommended comments.

Third, we note that Stack Overflow users often use code segments from open source projects as a part of the posted solutions. However, code segments from legacy or proprietary projects might not be shared at Stack Overflow.
Thus, our technique may fall short in recommending comments for the code segments from such projects. However, our findings clearly demonstrate its potential for open source projects. 

 
\section{Related Work}
\label{sec:related}
A number of studies from the literature generate comments for different granularities of code such as method \cite{methodaction,methodcomment,methodcontext}, class \cite{clssummary}, code segment \cite{csummarization, autocomment, methodaction}, and API method call \cite{codes, devcomm}.
\citeauthor{methodcomment} generate automatic comments both for an entire Java method \cite{methodcomment} and a group of code statements \cite{methodaction}.
They exploit semantics and linguistic clues in method signatures and identifier names, and syntactic code blocks (\eg\ conditional block, loop block) respectively, and generate equivalent natural language descriptions as the comments. 
Thus, their techniques are subject to the quality of identifier and method names, and they can generate comments only for a limited set of code structures (\eg\ one method body \cite{methodcomment}, groups of method calls \cite{methodaction}).
\citet{methodcontext} improve their technique \cite{methodcomment} by combining method context (\ie\ dependent methods) with method body.
\citet{autocomment} generate comments for a code segment of interest by mining developer's description on the similar posted code in Q \& A site answers.
Since they rely entirely on developer's description, quality of the generated comments is subject to the code documentation experience of the developer.
Moreover, no auto-generated comments \cite{methodcomment,methodaction,methodcontext} provide any insight into the \emph{quality, issues} or \emph{scopes} for further improvement of the code.
On the other hand, our heuristic-based technique mines various insights from Stack Overflow discussions led by a large technical crowd, refines them using NLP tools, and then 
recommends as code comments for a given code segment.

There exist other studies related to our work that mine \emph{API method descriptions} \cite{devcomm,codes,parnin,deficient} and \emph{code elements} (\eg\ method call) \cite{rigby} from developer communication documents such as bug reports and mailing lists, forums and programming Q \& A sites.
\citet{comprehend} conduct an observational study, and report that there exists a gap between program comprehension research and actual comprehension in practice.
While these studies focus on mining API documentations or comprehension patterns, we mine insightful comments from Stack Overflow for a given code segment that discuss issues, concerns or scopes for improvement in the code. 
From technical point of view, 
our approach complements the existing comment generation and API description mining techniques since they simply focus on explaining the code and the API elements respectively.

\section{Conclusion \& Future Work}
\label{sec:conclusion}
To summarize, we propose a mining technique that mines insightful comments from Stack Overflow for a given code segment, where the comments reveal identified \emph{issues, deficiencies} and \emph{scopes} for further improvement in the code.
We first conduct an exploratory study where we analyze Stack Overflow follow-up discussions, and report that the comments contain useful information for static analysis targeting software maintenance. 
We then propose a mining technique for automatic comment recommendation.
Our approach mines the comments by analyzing five aspects of comments-- \emph{popularity, relevance, comment rank, word count} and \emph{sentiment} expressed in the text. 
Experiments with 292 Stack Overflow code segments and 5,039 discussion comments show that our approach can extract the insightful comments with a promising \emph{recall} of 85.42\% and a MRR of 0.44 on average.
A user study with professional developers and 85 code segments also show that about 80\% of the recommended comments are found \emph{accurate, precise, concise} and \emph{useful} by the participants.
In future, we plan to 
integrate our technique in the IDE.

\balance


\bibliographystyle{plainnat}
\setlength{\bibsep}{0pt plus 0.3ex}
\scriptsize
\bibliography{sigproc}  
%
%
\end{document}